\documentclass{aa}  

\usepackage{natbib,twoopt}
\usepackage{float}
\usepackage{stfloats}
\usepackage[breaklinks=true,colorlinks=true, citecolor=blue, linkcolor=myrose, urlcolor=blue]{hyperref}
\bibpunct{(}{)}{;}{a}{}{,} 

\makeatletter
  \newcommandtwoopt{\citeads}[3][][]{\href{http://adsabs.harvard.edu/abs/#3}%
    {\def\hyper@linkstart##1##2{}%
     \let\hyper@linkend\@empty\citealp[#1][#2]{#3}}}
  \newcommandtwoopt{\citepads}[3][][]{\href{http://adsabs.harvard.edu/abs/#3}%
    {\def\hyper@linkstart##1##2{}%
     \let\hyper@linkend\@empty\citep[#1][#2]{#3}}}
  \newcommandtwoopt{\citetads}[3][][]{\href{http://adsabs.harvard.edu/abs/#3}%
    {\def\hyper@linkstart##1##2{}%
     \let\hyper@linkend\@empty\citet[#1][#2]{#3}}}
  \newcommandtwoopt{\citeyearads}[3][][]%
    {\href{http://adsabs.harvard.edu/abs/#3}
    {\def\hyper@linkstart##1##2{}%
     \let\hyper@linkend\@empty\citeyear[#1][#2]{#3}}}
\makeatother

\setcitestyle{authoryear,open={(},close={)}}
\usepackage{float}
\usepackage{multirow}
\usepackage{lscape}
\usepackage{xcolor}
\definecolor{myrose}{RGB}{200,0,100}

\usepackage{graphicx}
\usepackage{txfonts}
\usepackage{makecell}

\begin{document} 

    \title{Galaxy And Mass Assembly (GAMA): From filaments to voids, how extreme environment affects gas metallicity and SFR in galaxies}
    \titlerunning{Metallicity and SFR from filaments to voids}
    \authorrunning{J.A. Molina-Calzada}

    \subtitle{}

    \author{J. A. Molina-Calzada, \inst{1,} \inst{\,2}
            M. A. Lara-López, \inst{1,3}
            J. Gallego \inst{1,3}
            A. M. Hopkins \inst{5,6}
            Benne W. Holwerda \inst{4} \and
            A. R. López-Sánchez \inst{5,6,7}
            }

    \institute{Departamento de Física de la Tierra y Astrofísica. Universidad Complutense de Madrid (UCM), E-28040 Madrid, Spain
        \and
        Centro de Astrobiología, CSIC-INTA, Campus ESAC, Camino bajo del castillo s/n, E-28692 Villanueva de la Cañada, Madrid, Spain \\ \email{jmolina@cab.inta-csic.es}
        \and
        Instituto de Física de Partículas y del Cosmos, IPARCOS, Fac. C.C. Físicas, Universidad Complutense de Madrid, E-28040 Madrid, Spain
        \and
        Department of Physics, University of Louisville, Natural Science Building 102, 40292 KY, Louisville, USA
        \and
        School of Mathematical and Physical Sciences, Macquarie University, Sydney, NSW 2109, Australia
        \and
        Macquarie University Astrophysics and Space Technologies Research Centre, Sydney, NSW 2109, Australia
        \and
        ARC Centre of Excellence for All Sky Astrophysics in 3 Dimensions (ASTRO-3D), Australia}

    \date{Received 9 June 2025 / Accepted 7 July 2025}

 
    \abstract
    {The stellar mass$-$metallicity ($M_{\star}$--$Z$) and stellar mass$-$star formation rate ($M_{\star}$--SFR) relations are fundamental tools for understanding the evolution of star-forming (SF) galaxies. Combined with environmental factors, these relations provide valuable insights into how galaxies evolve.}
    {We analysed the $M_{\star}$--$Z$ and $M_{\star}$--SFR relations for SF galaxies, classified according to their environment, and compared them with a control sample of field galaxies. The aim was to quantify the differences in metallicity ($\Delta Z$) and star formation rate ($\Delta$SFR) among galaxies in different environments. To achieve this, we used data from the Galaxy And Mass Assembly (GAMA) survey, along with the filament catalogue that classifies galaxies into filaments, tendrils, and voids.}
    {The emission lines were corrected for dust extinction, SF galaxies were selected using the BPT diagram, and their $Z$ and SFR were estimated. Control samples were created for each type of environment, using field galaxies. The $M_{\star}$--$Z$ and $M_{\star}$--SFR relations were fitted using Bayesian and least-squares methods. The scaling relations for galaxies in different environments were compared to their corresponding control samples to establish robust differences.}
    {We determined that $\Delta Z$ increases as environments become denser. On the contrary, $\Delta$SFR increases as environments become less dense. Both results demonstrate significant differences between filaments and tendrils compared to voids. We also analysed galaxies in filaments and tendrils that do not belong to any group, and found little to no difference compared to their control sample. Morphology showed no significant deviation from the control sample.}
    {We find that galaxies in filaments and tendrils have higher metallicities and lower SFRs due to enriched environments, while void galaxies sustain high SFRs with low metallicities, likely driven by isolation and cold gas accretion. Our results indicate that local environmental factors, rather than large-scale structure, are the primary drivers of these differences.}

    
    \maketitle
%
\section{Introduction} \label{sec: Introduction}

Gas metallicity and star formation are among the key drivers of galaxy evolution. Detailed studies of these properties provide valuable insights into the role played by the environment.

Over the past decades, various empirical methods have been developed to estimate gas metallicities (Z) from galaxy emission lines, using different elements as tracers \citepads[e.g.]{2019MNRAS.487...79P}. Moreover, metallicity has been found to correlate with stellar mass, giving rise to the well-known mass$-$metallicity ($M_{\star}$--$Z$) relation. The study of this correlation has evolved and adapted to different galaxy types, providing valuable insights into its behaviour (\citeads{1979A&A....80..155L}; \citeads{2004ApJ...613..898T}; \citeads{2024A&A...682L..11S}, \citeyearads{2024A&A...681A.121S}; \citeads{2024AJ....168..226W}). 

Similarly, the star formation rate (SFR) measures the amount of stellar mass formed per unit time, and several empirical approaches exist for its estimation. One of the most widely used is based on the H$\alpha$ emission line. These methods have established a second key correlation, the mass$-$star formation rate ($M_{\star}$--SFR) relation, which has proven robust across various contexts (\citeads{1997ApJ...489..559G}; \citeads{2008A&A...483..107B}; \citeads{2016MNRAS.461..458D}; \citeads{2021MNRAS.505..540T}). Together, these relations reveal that more massive galaxies generally exhibit higher $Z$ and SFR values, while lower-mass systems show lower values for both parameters. Furthermore, studies have proposed a combined relation involving stellar mass, metallicity, and SFR, in order to provide a more comprehensive view of galaxy evolution (\citeads[e.g.]{2010A&A...521L..53L}, \citeyearads{2013MNRAS.434..451L}). 

These correlations, however, have also been explored in broader contexts. One key aspect is morphology: the SFR appears to be influenced by galaxy structure \citepads{2022MNRAS.515.3875P}, with spiral disks following distinct trends \citepads{2015MNRAS.449..820W}, while elliptical galaxies tend to exhibit higher metallicities (\citeads{2006astro.ph.10831P}; \citeads{2011A&A...525A..61P}). Morphology itself has been linked to the galactic environment, with evidence suggesting that morphological type plays a more significant role than environment in determining parameter evolution (\citeads[e.g.]{2018MNRAS.481.3456C}; \citeads{2024ApJ...964L..33L}). Nevertheless, environment remains a crucial factor, as shown by \citeads{2024A&A...686A..40S}, who found that galaxy mergers are more frequent in underdense regions, and impact various galactic properties. Finally, the temporal evolution of these parameters has also been widely investigated (\citeads[e.g.]{2007ApJ...660L..43N}; \citeads{2021ApJ...919..143H}), particularly in the era of next-generation telescopes like JWST, which has revealed that at high redshifts galaxy evolution remains strongly environment-dependent (\citeads[e.g.]{2024A&A...684A..75C}; \citeads{2024ApJ...964L..33L}). Lastly, it is important to highlight the current role of machine learning in the study of these parameters. Some machine learning studies are capable of deriving $Z$ and SFR from galaxy images (\citeads[e.g.]{2019MNRAS.484.4683W}; \citeads{2024ApJ...967..152A}).

Building on studies of environmental dependence in $Z$ and SFR, several authors have examined these relationships further. \citetads{2021MNRAS.508.1817S} found that local density, rather than group-centric distance, more strongly affects both SFR and $Z$, indicating that stellar mass is likely the main factor driving quenching in star-forming galaxies. Similarly, \citetads{2012MNRAS.423.3679W} argued that stellar mass primarily influences SFRs, with minimal environmental effects in star-forming galaxies. They also suggested that SFR-density trends are mainly driven by changes in galaxy morphology. On the other hand, \citetads{2008AJ....135.1877E} showed that for galaxies in pairs, the bulge fraction significantly affects $Z$. Finally, the quenching process described by \citetads{2021MNRAS.508.1817S}, where red galaxies are often found in denser environments, has been linked to both $Z$ and SFR in other studies, such as \citetads{2014MNRAS.445.2125M}. However, some reports suggest that processes such as starvation or ram-pressure stripping may play a more prominent role than quenching \citepads[e.g.]{2020MNRAS.491.5406T}, with the latter being one of the main mechanisms responsible for suppressing the inflow of metal-poor gas, thereby preventing dilution \citepads{2025MNRAS.540L..58K}.

In addition to these parameters, large-scale structure has also been studied. Many papers, such as those reported by \citetads{1986ApJ...302L...1D} and \citetads{2014MNRAS.440L.106A}, have paved the way for classifying the distribution of galaxies in the Universe. Galaxies tend to cluster into groups and clusters ($\sim$ 1 Mpc) that gravitationally clump together to form super-clusters ($\sim$ 10 Mpc). These super-clusters appear to be linked together in the form of chains of galaxies, creating filaments ($\sim$ 100 Mpc) that surround voids, which correspond to low-density regions of the Universe. Attached to these filaments are branches that protrude from these structures, known as ‘tendrils’. These regions were first introduced by \citetads{2014MNRAS.440L.106A}. Based on these previous reports, galaxies can be classified into the following groups: galaxies in filaments, tendrils galaxies, and void galaxies.

\citetads{2024MNRAS.534.1682O} concludes that the differences in star formation activity and morphology between galaxies in filaments and those in the field are primarily driven by local density, rather than the filaments themselves. Focusing on galaxies in voids, \citetads{2023MNRAS.524.5768P} and  \citetads{2024arXiv241102129A} argue that these galaxies are predominantly late type, with a morphology that is independent of both void size and galaxy density. In addition to filaments and voids, \citetads{2023A&A...680A.111D} also consider walls (dense regions of galaxies and dark matter that form flat or filament-like structures, connecting galaxy clusters and filaments). They all show that void galaxies tend to have slightly lower stellar metallicities compared to those in filaments and walls, and significantly lower than cluster galaxies. In addition, these differences are most evident in low-mass, extended-SFH, spiral, and blue galaxies. Despite these findings, many authors (\citeads{2022A&A...668A..69E}; \citeads{2024MNRAS.528.4139H}; \citeads{2024MNRAS.534.1682O}; \citeads{2024eas..conf.1965T}) emphasise that both local and large-scale environments primarily drive galaxy evolution. High-density regions lead to quenched star formation and early-type galaxies, while low-density regions show more varied star formation histories. Although some authors give greater weight to the local environment (\citeads{2004MNRAS.353..713K}; \citeads{2012MNRAS.420.1481V}; \citeads{2013MNRAS.431..167R}).

The goal of this work is to examine the relation between star formation and the metallicity of galaxies and their position in this large-scale structure. To this end, we used a well-determined sample with spectroscopic redshifts and ample ancillary data. 

This paper is structured as follows. Section \ref{sec: Sample selection} explains the Galaxy And Mass Assembly (GAMA) survey and how the filtering process was carried out to select the sample of galaxies studied here. The fluxes of these filtered galaxies were corrected for dust extinction, and star-forming (SF) galaxies were selected using the Baldwin-Phillips-Terlevich (BPT) diagram. Metallicity and SFR were then derived from their emission lines. Afterwards, the different types of galaxies (filaments, tendrils, and voids) are explained, and control samples are generated from field galaxies of GAMA under specific established conditions and tests. In Section \ref{sec: Results}, the scaling relations found from the different samples are analysed and compared. For this purpose, a classical and Bayesian analysis was conducted to compare the differences $\Delta Z$ and $\Delta$SFR of the galaxies in the various environments with a control sample. Finally, Section \ref{sec: Discussion} discusses the results obtained, and we present the conclusions of this work in Section \ref{sec: Conclusions}. 

\section{Sample selection} \label{sec: Sample selection}

\subsection{GAMA survey} \label{sec: GAMA survey}

The Galaxy and Mass Assembly (GAMA) (\citeads{2011MNRAS.413..971D}; \citeads{2015MNRAS.452.2087L}) is a large survey whose primary objective is to map the distribution of galaxies at various scales, with a focus on obtaining high-quality spectroscopic data for each galaxy. The data were taken with the 3.9\,m Anglo-Australian Telescope (AAT) employing the AAOmega multi-object spectrograph \citepads{2006SPIE.6269E..0GS} and the 2dF fibre feed. The spectra were taken using a 2 arcsec diameter fibre, covering a range between 3\,740\text{\AA} to 8\,850\text{\AA}, with a resolution of 3.2\text{\AA}. The survey covers a large area of the sky, approximately 286 square degrees, and includes over  340\,000 galaxies. This study focuses on the GAMA Phase 1, which collects 192\,545 galaxies in 48 square degrees, and a limiting Petrosian magnitude of m$_r<$19.8\,mag in one field and m$_r<$19.4\,mag in the other two (for further details see \citetads{2010MNRAS.404...86B}, \citetads{2011MNRAS.413..971D}, \citetads{2013MNRAS.430.2047H}, and \citetads{2015MNRAS.452.2087L}).

On the other hand, the emission lines from GAMA were measured using the  Gas AND Absorption Line Fitting (GANDALF) algorithm \citepads{2006MNRAS.366.1151S}, which is a simultaneous emission and absorption line fitting algorithm. GANDALF measures and corrects for stellar absorption of the emission lines. All these line measurements include the emission line flux, the equivalent width (EW), and the signal-to-noise ratio (S/N) for the most prominent emission lines, among other parameters (for more details, see \citetads{2018MNRAS.475.4223G}).

The full GAMA catalogue is matched with two other catalogues, one for stellar masses \citepads{2011MNRAS.418.1587T}, and one for filaments \citepads{2014MNRAS.438..177A}, both from the GAMA survey. The \citetads{2011MNRAS.418.1587T} catalogue estimates the stellar mass-to-light ratio ($M_{\star}$/L) from optical photometry using stellar population synthesis based on the \citetads{2003MNRAS.344.1000B} model. The \citetads{2014MNRAS.438..177A} catalogue uses an adapted minimal spanning tree protocol called Friend of Friend (FoF) algorithm \citepads{2013MNRAS.429..556M}, which groups galaxies based on their proximity to one another, thus creating a network of connected galaxies. It identifies galaxies within a specified linking length and classifies them based on their connection to nearby galaxies. 

The spectroscopic sample of galaxies extracted from the GAMA  catalogue used here has 192\,545 galaxies. First, we filtered the data based on  signal-to-noise ratio (S/N) and redshift ($z$) thresholds to ensure the reliability of our results, as detailed below:

\begin{itemize}

    \item S/N\,$>$\,3 for H$\alpha$ and [O\,\textsc{iii}]\,$\lambda$5007, and S/N\,$>$\,2 for H$\beta$ and [N\,\textsc{ii}]\,$\lambda$6583.

    \item $z$\,$<$\,0.3 to ensure the H$\alpha$ emission line remains within the optical range, as it shifts out at higher redshifts.
    
\end{itemize}

Table \ref{tab: filter_table} summarises the number of galaxies selected after successively applying the filtering conditions.

\begin{table}[H]
    \caption{Galaxies selected from GAMA after the S/N and redshift cuts.}
    \centering
    \begin{tabular}{c c}
        \hline\hline
        \makecell{Condition} & \makecell{Number of galaxies} \\   
        \hline
        \makecell{Initial sample} & 192\,545 \\
        \makecell{After S/N cut (i)} & 18\,247 \\ 
        \makecell{After redshift cut (ii)} & 17\,283 \\ 
        \hline
    \end{tabular}
    \tablefoot{The redshift cut shows the total number of galaxies selected due to both conditions explained.}
    \label{tab: filter_table}
\end{table}

Therefore, the initial number of galaxies available in the GAMA catalogue for this study was reduced from 192\,545 to 17\,283 after the filtering process.

Next, we corrected all the emission lines for dust extinction. The extinction curve used in this work is from \citetads{1989ApJ...345..245C}, we assume a case B of recombination \citepads{1989agna.book.....O} and $H\alpha/H\beta$ = 2.86. Therefore, the corrected flux is given by $\mathrm{F}(\lambda)_{corr}=\mathrm{F}(\lambda)_{obs}\cdot 10^{0.4\mathrm{A}_\lambda}$, where $\mathrm{F}(\lambda)_{corr}$ and $\mathrm{F}(\lambda)_{obs}$ correspond to corrected and observed flux, respectively. The parameter $\mathrm{A}_{\lambda}$ is the extinction for a given wavelength. When $H\alpha/H\beta$ is lower than the theoretical value, then the adopted correction is set to zero.

\subsection{BPT diagram} \label{sec: BPT Diagram}

The  Baldwin-Phillips-Terlevich (BPT) diagram (\citeads{1981PASP...93....5B}; \citeads{1987ApJS...63..295V}) is commonly used to classify and understand the ionisation sources of emission-line galaxies. This diagram allows us to classify the galaxies from the GAMA survey into three groups: star-forming (SF), active galactic nucleus (AGN), and composite. 

\begin{figure}[ht]
    \centering
    \includegraphics[width=\hsize]{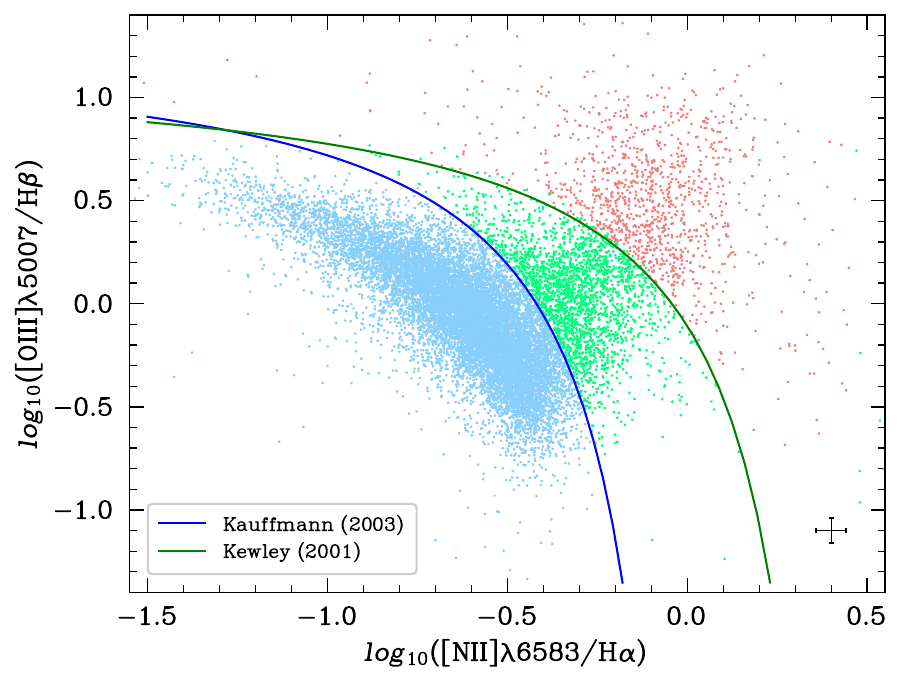}
    \caption{BPT diagram for GAMA galaxies. Galaxies below the \citetads{2003MNRAS.346.1055K} function correspond to SF galaxies (blue points), while the galaxies above the \citetads{2001ApJ...556..121K} function correspond to AGN galaxies (red points). Galaxies that fall between the \citetads{2001ApJ...556..121K} and \citetads{2003MNRAS.346.1055K} functions are classified as composite galaxies (green points). The error cross corresponds to the median errors of the axes calculated using Monte Carlo propagation.}
    \label{fig: BPT_Diagram}
\end{figure}

The BPT diagram in Figure \ref{fig: BPT_Diagram} is plotted using the emission lines of [O\,\textsc{iii}]\,$\lambda$5007, [N\,\textsc{ii}]\,$\lambda$6583, H$\alpha$ and H$\beta$ from the 17\,283 galaxies selected. We used the curve of \citetads{2003MNRAS.346.1055K} to separate composite from AGN galaxies, while the curve of \citetads{2001ApJ...556..121K} separates composite from SF galaxies.

The estimation of the error bars, shown in Figure \ref{fig: BPT_Diagram}, was done using Monte Carlo propagation. This method generates simulated data for each variable from a multivariate t-distribution, using the means and the full covariance matrix, which includes the diagonal elements (variances) and the off-diagonal terms (covariances) derived from the uncertainties of the original variables.

Based on Figure \ref{fig: BPT_Diagram}, the different types of galaxies are classified in the BPT diagram as shown in Table \ref{tab: galaxy_classification}.

\begin{table}[H]
    \caption{Number of galaxies of each type based on the BPT diagram.}
    \centering
    \begin{tabular}{c c}
        \hline\hline
        \makecell{Galaxy type} & \makecell{Number of galaxies} \\   
        \hline
        \makecell{SF } & 13\,689 \\
        \makecell{Composite } & 2\,336 \\
        \makecell{AGN } & 1\,258 \\
        \hline
        \hline
    \end{tabular}
    \label{tab: galaxy_classification}
\end{table}

The data in Table \ref{tab: galaxy_classification} indicate that a total of 79\% of the sample (13\,689 galaxies) consists of SF galaxies, which is the focus of our analysis hereafter.

\subsection{Metallicity and star formation rate estimation} \label{sec: Metallicity and star formation rate}

Gas metallicities are estimated using the extinction-corrected fluxes of the SF galaxies, and the calibration method from \citetads{2004MNRAS.348L..59P}. 

\begin{equation}
    12 + \log_{10}(\mathrm{O}/\mathrm{H}) = 8.73 - 0.32 \cdot \mathrm{O3N2},
    \label{eq: Pattini_metallicity}
\end{equation}

where the parameter O3N2 is defined as

\begin{equation}
    \mathrm{O3N2} = \log_{10}\left(\frac{\left[\mathrm{O}\,\textsc{iii}\right]\,\lambda5007/\mathrm{H}\beta}{\left[\mathrm{N}\,\textsc{ii}\right]\,\lambda6583/\mathrm{H}\alpha}\right),
    \label{eq: O3N2_parameter}
\end{equation}

and is in the range -1 $<$ $\mathrm{O3N2}$ $<$ 1.9. We adopted the O3N2 calibration as it provides reliable metallicity estimates for large samples of star-forming galaxies \citepads{2008ApJ...681.1183K}. Additionally, since the emission-line ratios involved are close in wavelength, this method is minimally affected by dust extinction \citepads{ 2013MNRAS.434..451L}. 

On the other hand, we use the relation of \citetads{1998ARA&A..36..189K} to estimate the SFR. This relation is expressed as

\begin{equation}
    \mathrm{SFR}_{\mathrm{Hopkins}} [\mathrm{M}_\odot/\text{yr}]=\frac{\mathrm{L}_{\mathrm{H} \alpha}}{1.27 \times 10^{34}\,W},
\end{equation}

where the luminosity of the H$\alpha$ line, corrected for dust obscuration, is defined by \citetads{2003ApJ...599..971H} as

\begin{equation}
    \begin{aligned}
    \mathrm{L}_{\mathrm{H} \alpha}= & \mathrm{EW}_{\mathrm{H} \alpha} \times 10^{-0.4\left(M_r-34.10\right)} \times \frac{3 \times 10^{18}}{[6564.61(1+z)]^2} \\
    &\times \left(\frac{F(\mathrm{H} \alpha)_{\mathrm{obs}} / F(\mathrm{H} \beta)_{\mathrm{obs}}}{2.86}\right)^{2.36},
    \label{eq: Hopkins_calibration}
    \end{aligned}
\end{equation}

where $\mathrm{F}(\mathrm{H} \alpha)_{\mathrm{obs}}$ and $\mathrm{F}(\mathrm{H} \beta)_{\mathrm{obs}}$ are the observed fluxes of the H$\alpha$ and H$\beta$ lines, respectively, corrected by stellar absorption by GANDALF. The EW$_{\mathrm{H} \alpha}$ denotes the equivalent width of the H$\alpha$ line in angstroms. Finally, $M_r$ denotes the absolute Petrosian magnitude in the r-band. With these considerations, the representation of the $M_{\star}$--$Z$ and $M_{\star}$--SFR relations can be established.

It is important to note that the fibre size of the AAT could introduce some bias in the estimation of $Z$ and SFR parameters. This issue is particularly relevant for SFR measurements due to the galaxy's angular size. If a galaxy is large, its outer regions might not fit entirely within the fibre, potentially leading to incomplete SFR measurements. However, this potential bias is mitigated by the calibration method used by \citetads{2003ApJ...599..971H}, which accounts for such discrepancies in Equation \ref{eq: Hopkins_calibration}. In contrast, the $Z$ parameter is less significantly affected. According to \citetads{2005PASP..117..227K}, as long as at least 20\% of the light of the galaxy falls within the optical fibre, the gas metallicity provides a reasonable approximation of the total value.

\subsection{Galaxies in filaments, tendrils, and voids} \label{sec: Galaxies in filaments, tendrils, and voids}

The distribution of galaxies in the Universe is not homogeneous. At large scales ($\sim$100\,Mpc), the galaxies are arranged into vast, elongated structures known as filaments, from which smaller branches called tendrils emanate \citepads{2014MNRAS.440L.106A}. The regions between these filaments are characterised by a lack of galaxies and are referred to as voids. 

For this work, we used the classification provided by \citetads{2014MNRAS.438..177A}, who applied the FoF algorithm \citepads{2013MNRAS.429..556M} to classify the galaxies in the GAMA survey. This classification divides galaxies into the following types.

\begin{itemize}

    \item Galaxies in filaments. This classification includes all galaxies within an orthogonal distance of 4.12$\,h^{-1}$\,Mpc from filaments. Therefore, they are situated in dense environments and are susceptible to the influences of their surroundings.
    
    \item Tendril galaxies. Included in this category are all galaxies within an orthogonal distance (4.56$\,h^{-1}$\,Mpc) from tendrils, and simultaneously located beyond a minimum distance of 4.12$\,h^{-1}$,Mpc from any filament. Tendrils are slender, thread-like extensions that serve as connecting links between certain filaments. This term was first introduced by \citetads{2014MNRAS.440L.106A}.
    
    \item Void galaxies. These are galaxies that are more than 4.56$\,h^{-1}$\,Mpc from both filament and tendril structures. As a result, they are situated in largely underdense regions called voids, where they remain unaffected by external influences.
    
\end{itemize}

Considering these different types of galaxies, the sample of SF galaxies used in this study was classified into these three types, by matching the GAMA catalogue and the catalogue of filaments \citepads{2014MNRAS.438..177A}, as shown in Table \ref{tab: type_galaxies}.

\begin{table}[H]
    \caption{Classification of the sample of SF galaxies considered in this study according to the different environments.}                   
    \centering                          
    \begin{tabular}{c c}       
        \hline\hline                
        \makecell{Environment} & \makecell{Number of SF galaxies} \\   
        \hline                        
        Filaments & 2\,068 \\      
        Tendrils & 1\,466 \\      
        Voids & 189 \\         
        \hline 
    \end{tabular}
    \label{tab: type_galaxies}
\end{table}

From the initial 13\,689 SF galaxies, 15\% correspond to galaxies in filaments, 11\% to tendril galaxies, and 1\% to void galaxies. The percentage of the remaining unclassified galaxies corresponds mostly to field galaxies and will be used in the following section to create the control sample.

\subsection{Control samples} \label{sec: Generation of control samples}

\begin{table*}[ht]
    \caption{p-values for the Kolmogorov-Smirnov (KS) and Anderson-Darling (AD) tests on stellar mass and colour.}                 
    \centering
    \small
    \begin{tabular}{c | c c c c c c c c c c}       
        \hline\hline                
        Environment & \multicolumn{2}{c}{Filaments} & \multicolumn{2}{c}{Fil. non-grpd.} & \multicolumn{2}{c}{Tendrils} & \multicolumn{2}{c}{Ten. non-grpd.} & \multicolumn{2}{c}{Voids} \\   
        \hline
         Parameter & Stellar mass & Colour & Stellar mass & Colour & Stellar mass & Colour & Stellar mass & Colour & Stellar mass & Colour \\   
        \hline                        
        KS test & 0.18 & 0.56 & 0.43 & 0.55 & 0.14 & 0.11 & 0.12 & 0.73 & 0.81 & 0.48 \\      
        AD test & 0.05 & 0.20 & $>$\,0.25 & $>$\,0.25 & 0.06 & 0.20 & 0.22 & $>$\,0.25 & $>$\,0.25 & $>$\,0.25 \\ 
        \hline                                 
    \end{tabular}
    \tablefoot{The tests are applied to the Empirical Cumulative Distribution Functions (ECDFs) of the control samples and the problem samples of galaxies in filaments, tendrils, and voids.}
    \label{tab: p-values-combined}
\end{table*}

In this study, a control sample refers to a subset of galaxies selected as a reference to compare the properties of galaxies situated in filaments, tendrils, and voids. This sample is carefully maintained under specific conditions to ensure its independence from the large-scale structures under analysis, thereby providing a reliable baseline for comparison. 

We constructed such a control sample using field galaxies, selected to serve as a reference population. Field galaxies are identified from the original GAMA spectroscopic catalogue by excluding all galaxies in filaments, tendrils, and voids. In addition, we also removed galaxies that are members of groups or pairs. These exclusions are based on classifications from the GAMA filament catalogue \citepads{2014MNRAS.438..177A}, and the GAMA Galaxy Group Catalogue (G3C) \citepads{2011MNRAS.416.2640R}. We note that we worked only with SF galaxies with the same S/N and redshift cuts described previously in this section. By applying these criteria, we obtained a well-defined and uncontaminated sample of field galaxies for our analysis. 

Previous studies investigating environmental effects have generated control samples using similar methodologies and criteria (\citeads[e.g.]{2008AJ....135.1877E}; \citeads{2012MNRAS.426..549S}; \citeads{2020RAA....20...20C}; \citeads{2021MNRAS.501.2969G}; \citeads{2021MNRAS.508.1817S}). We note that field galaxies are used as a control sample because they are not exposed to extreme environments, that is, they are neither located in structures such as filaments or tendrils, nor highly underdense regions such as voids. This makes them suitable for defining control samples, as done in previous studies \citepads[e.g.]{2008AJ....135.1877E}.

To ensure the validity of our study, these field galaxies must share similar characteristics with those in filaments, tendrils, and voids. We followed previous works and used stellar mass and colour as our main criteria (\citeads[e.g.]{2008AJ....135.1877E}, \citeyearads{2011MNRAS.418.2043E}, \citeyearads{2013MNRAS.435.3627E}), since these two parameters robustly characterise the main properties of a galaxy (\citeads{2004ApJ...600..681B}; \citeads{2010ApJ...721..193P}). Given that our sample of galaxies is relatively local (z\,$<$\,0.3), the evolutionary effects are negligible \citepads[e.g.]{2009A&A...505..529L}. Therefore, we ensure that the differences in stellar mass and colour between galaxies in filaments, tendrils, and voids and those in the control sample remain minimal, adhering to specific selection criteria.

These selection criteria, based on stellar mass and colour, were established using the Kolmogorov-Smirnov (KS) and Anderson-Darling (AD) tests as references. These tests assess the null hypothesis that two samples originate from the same population without requiring the specification of the distribution function of that population. The KS test is good enough to verify whether the created control sample and the sample of galaxies in filaments, tendrils, or voids can be assumed to be drawn from the same galaxy population. However, we complement this test with the AD test since it is more sensitive to the tails of the Empirical Cumulative Distribution Function (ECDF), thus ensuring that the samples are similar. The KS and AD tests accept the null hypothesis that both samples can be assumed to be drawn from the same population if the p-value is higher than the significance level $\alpha$ = 0.05.

During the creation of the control sample, additional considerations include limiting the number of matched field galaxies per target galaxy. For galaxies in filaments, a maximum of 20 field galaxies per filament galaxy was set. For tendril galaxies, this limit ranges from 2 to 20, while for void galaxies, it varies between 3 and 8. Similar studies conducted using the SDSS catalogue have set this limit at 20 \citepads[e.g.]{2008AJ....135.1877E}, but in our case, since the GAMA sample size is smaller,  it does not allow for the creation of such large control samples.

\begin{figure}[ht]
    \centering
    \includegraphics[width=\hsize]{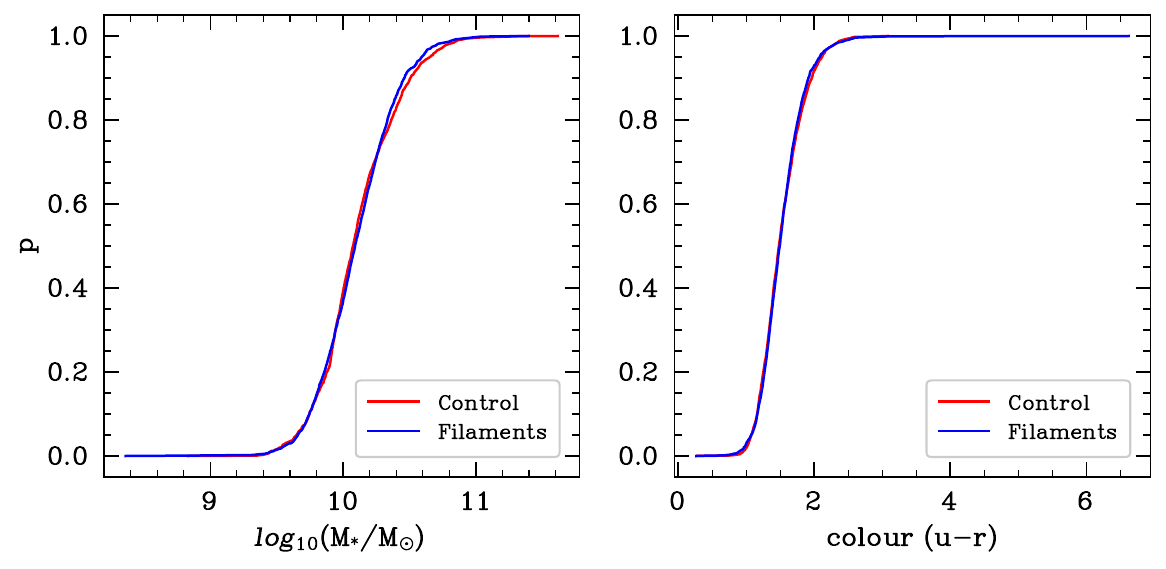}
    \includegraphics[width=\hsize]{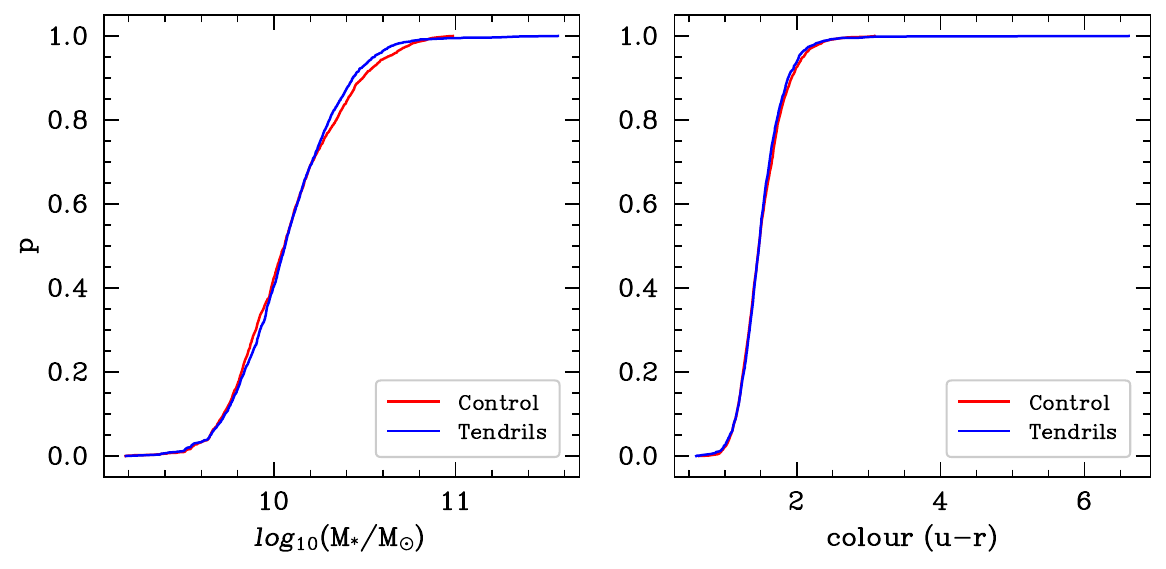}
    \includegraphics[width=\hsize]{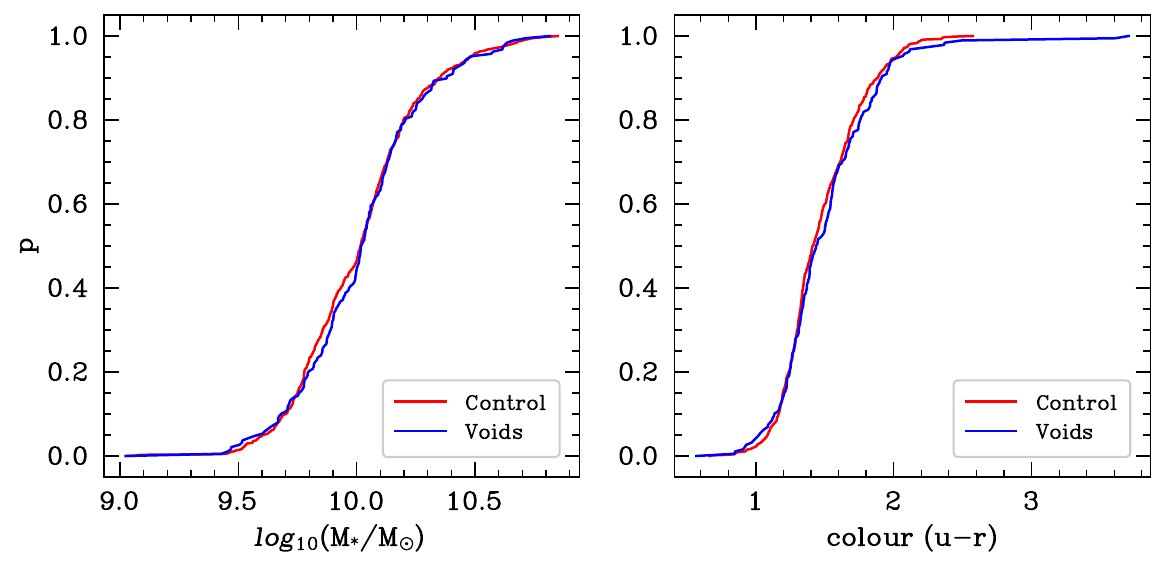}
    \caption{From top to bottom, the Empirical Cumulative Distribution Function (ECDF) of the control sample of galaxies in filaments, tendrils and voids for stellar mass (left) and colour (right). The red line shows the control sample, and the blue line the problem sample.}
    \label{fig: ECDF_filaments}
\end{figure}

These criteria allow us to obtain a control sample of 2\,114 field galaxies versus the 2\,068 galaxies in filaments. The ECDFs for the stellar mass and colour of both the control sample and galaxies in filaments are shown in Figure \ref{fig: ECDF_filaments}. Furthermore, since all the p-values collected in Table \ref{tab: p-values-combined} are greater than the significance level, $\alpha$ = 0.05, it can be assumed that the control sample for galaxies in filaments and the sample of galaxies in filaments follow the same distribution. 

Similarly, for tendril galaxies, the control sample includes 2\,181 field galaxies versus 1\,466  galaxies in tendrils. The corresponding ECDFs and p-values, derived from the KS and AD tests, are also included in Figure \ref{fig: ECDF_filaments} and Table \ref{tab: p-values-combined}, respectively.

Finally, for void galaxies, the control sample includes 704 field galaxies compared to 189 void galaxies. The ECDFs and p-values for this case are presented in the same figures and tables. These results confirm that the control samples are representative of their respective galaxy environments.  

In addition, it is important to highlight that the KS and AD tests are designed for one-dimensional samples, and their extension to higher dimensions can lead to issues. Thus, we tested our control samples by applying the two-dimensional KS test using the \texttt{mks$\_$test} package from Python \citep{naaman2021tight}. On the other hand, note that the upper limit of $>$\,0.25 for some of the p-values obtained from the AD test is due to the limitations of the Python package used (\texttt{scipy} package; \citeads{2020NatMe..17..261V}).

\section{Results} \label{sec: Results}

Once the control samples have been determined for each sample of galaxies (filaments, tendrils, and voids), the $M_{\star}$--$Z$ and $M_{\star}$--SFR relations can be studied. Specifically, the goal is to establish if there are differences in metallicity, $\Delta Z$, or star formation rate, $\Delta$SFR, between the galaxy samples and their corresponding control samples. To quantify these differences, we will fit the relations using two methods: least-squares fitting (Sect. \ref{sec: Least-squares method}) and  Bayesian fitting (Sect. \ref{sec: Bayesian method}).

Our fitting approach involves modelling the $M_{\star}$--$Z$ and $M_{\star}$--SFR relations for the corresponding control samples first. Then, for the filament, tendril, and void galaxy samples, we fix the higher-order coefficients of their corresponding control samples and fit only the zero-point.
In this way, only the intercept (zero-order coefficient) is adjusted for each environment-specific sample. The quantities $\Delta$SFR and $\Delta Z$ are defined as the differences between the fitted zero-points of these samples and those of their respective control samples.

\begin{table*}[t]
    \centering
    \caption{Differences in $Z$ and SFR between galaxies in different environments and their control samples, using the least-squares method.}
    \begin{tabular}{c c c c c l l l}
    \hline\hline
    $\Delta Z_{filaments}$ & $\Delta Z_{fil. non-grpd.}$ & $\Delta Z_{tendrils}$ & $\Delta Z_{ten. non-grpd}$ & $\Delta Z_{voids}$ & \multicolumn{3}{c}{} \\ 
    \hline
    (0.020 $\pm$ 0.003)\,dex &($-$0.0035 $\pm$ 0.0038)\,dex & (0.018 $\pm$ 0.003)\,dex & (0.0026 $\pm$ 0.0037)\,dex & ($-$0.014 $\pm$ 0.008)\,dex & \multicolumn{3}{c}{} \\ 
    \hline\hline
    $\Delta \mathrm{SFR}_{filaments}$ & $\Delta \mathrm{SFR}_{fil. non-grpd.}$ & $\Delta \mathrm{SFR}_{tendrils}$ & $\Delta \mathrm{SFR}_{ten. non-grpd.}$ & $\Delta \mathrm{SFR}_{voids}$ & \multicolumn{3}{c}{} \\ 
    \hline
    (0.024 $\pm$ 0.015)\,dex &($-$0.0058 $\pm$ 0.0191)\,dex & (0.031 $\pm$ 0.016)\,dex & (0.0055 $\pm$ 0.0175)\,dex & (0.078 $\pm$ 0.036)\,dex & \multicolumn{3}{c}{} \\ 
    \hline
    \end{tabular}
    \tablefoot{The associated errors are estimated via bootstraping. The environments considered are galaxies in filaments (grouped and non-grouped), tendrils (grouped and non-grouped), and voids.}
    \label{tab: Relation_coefLS}
\end{table*}

\begin{table*}[t]
    \centering
    \caption{Differences in $Z$ and SFR between galaxies in different environments and their control samples, using the Bayesian method.}
    \begin{tabular}{c c c c c l l l}
    \hline\hline
    $\Delta Z_{filaments}$ & $\Delta Z_{fil. non-grpd.}$ & $\Delta Z_{tendrils}$ & $\Delta Z_{ten. non-grpd}$ & $\Delta Z_{voids}$ & \multicolumn{3}{c}{} \\ 
    \hline
    (0.019 $\pm$ 0.003)\,dex & ($-$0.0040 $\pm$ 0.0038)\,dex & (0.016 $\pm$ 0.003)\,dex & (0.0017 $\pm$ 0.0037)\,dex & ($-$0.017 $\pm$ 0.008)\,dex & \multicolumn{3}{c}{} \\ 
    \hline\hline
    $\Delta \mathrm{SFR}_{filaments}$ & $\Delta \mathrm{SFR}_{fil. non-grpd.}$ & $\Delta \mathrm{SFR}_{tendrils}$ & $\Delta \mathrm{SFR}_{ten. non-grpd.}$ & $\Delta \mathrm{SFR}_{voids}$ & \multicolumn{3}{c}{} \\ 
    \hline
    (0.041 $\pm$ 0.016)\,dex & (0.0053 $\pm$ 0.0191)\,dex & (0.061 $\pm$ 0.017)\,dex & (0.028 $\pm$ 0.019)\,dex & (0.092 $\pm$ 0.040)\,dex & \multicolumn{3}{c}{} \\ 
    \hline
    \end{tabular}
    \tablefoot{The associated errors correspond to the standard deviation of the Markov chains. The environments considered are galaxies in filaments (grouped and non-grouped), tendrils (grouped and non-grouped), and voids.}
    \label{tab: Relation_coefB}
\end{table*}

\subsection{Least-squares method} \label{sec: Least-squares method}

This method was implemented using the \texttt{lmfit} package from Python \citepads{2016ascl.soft06014N}, incorporating weights defined as the inverse of the squared residuals. This approach ensures that galaxies with lower residuals contribute more significantly to the fit, while those with higher residuals have a reduced impact.

Figure \ref{fig: Combined_Plot_LS} presents fits for the $M_{\star}$--$Z$ and $M_{\star}$--SFR relations using the least-squares method. A third-order polynomial fit was used for the $M_{\star}$--$Z$ relation, while a linear regression was applied to the $M_{\star}$--SFR relation. Additionally, the residuals of the fit approach zero, indicating a robust and well-constrained fit.

For the $M_{\star}$--$Z$ relations, the $\Delta Z$ differences in metallicity between the galaxies of each type and their corresponding control samples are small. However, for the $M_{\star}$--SFR relations, the $\Delta$SFR differences in SFR become appreciable (see Figure \ref{fig: Combined_Plot_LS}). To robustly analyse these results and draw conclusions, it is essential to consider the associated errors of these differences. Table \ref{tab: Relation_coefLS} summarises the differences in $Z$ and SFR found and their associated errors calculated using bootstrapping. 

\subsection{Bayesian method} \label{sec: Bayesian method}

Bayesian regression is a type of conditional modelling where a linear combination of other variables describes the mean of a variable. The goal is to obtain the posterior probability distributions of the regression coefficients. This method was implemented here by creating models in \texttt{Stan} \citepads{2017JSS....76....1C} using the R programming language. Error estimates are accounted for in these regression models. Therefore, using a Markov Chain Monte Carlo (MCMC) convergence diagnostic makes it possible to obtain robust fits for the scaling relations.

\begin{figure*}
    \centering
    \includegraphics[width=0.93\hsize]{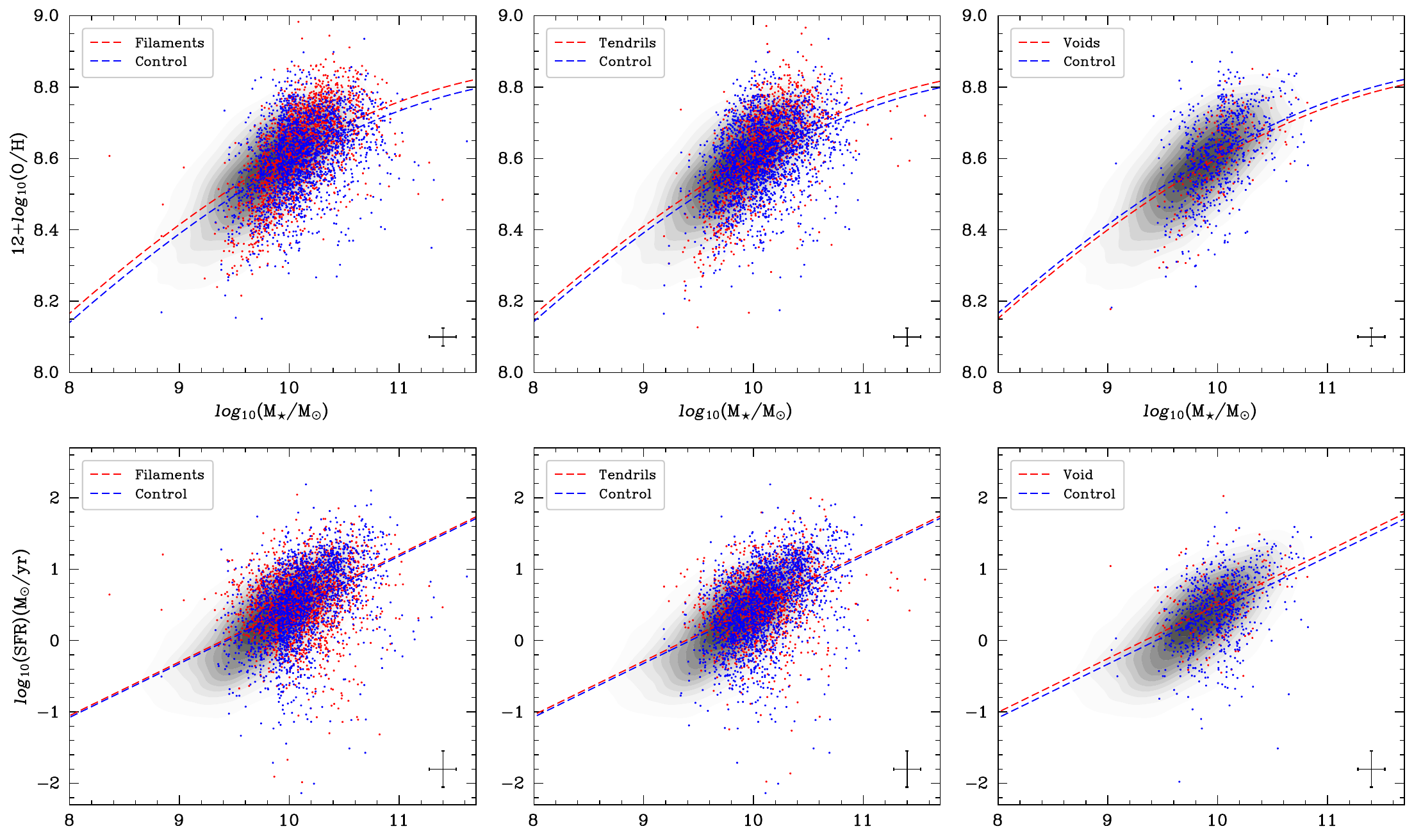}
    \caption{$M_{\star}$--$Z$ (top) and $M_{\star}$--SFR (bottom) relations, using the \citetads{2004MNRAS.348L..59P} and \citetads{2003ApJ...599..971H} calibrations respectively. The galaxies in filaments (left), tendril galaxies (centre), and voids (right) are marked as red points alongside their corresponding control samples in blue. All SF galaxies are in the background as a black density scatter plot. The dashed lines represent the linear regressions calculated by the least-squares method for each galaxy type (in red) and its corresponding control sample (in blue). Error bars are the median errors calculated using Monte Carlo propagation.}
    \label{fig: Combined_Plot_LS}

    \vspace{0.4cm}
    
    \includegraphics[width=0.93\hsize]{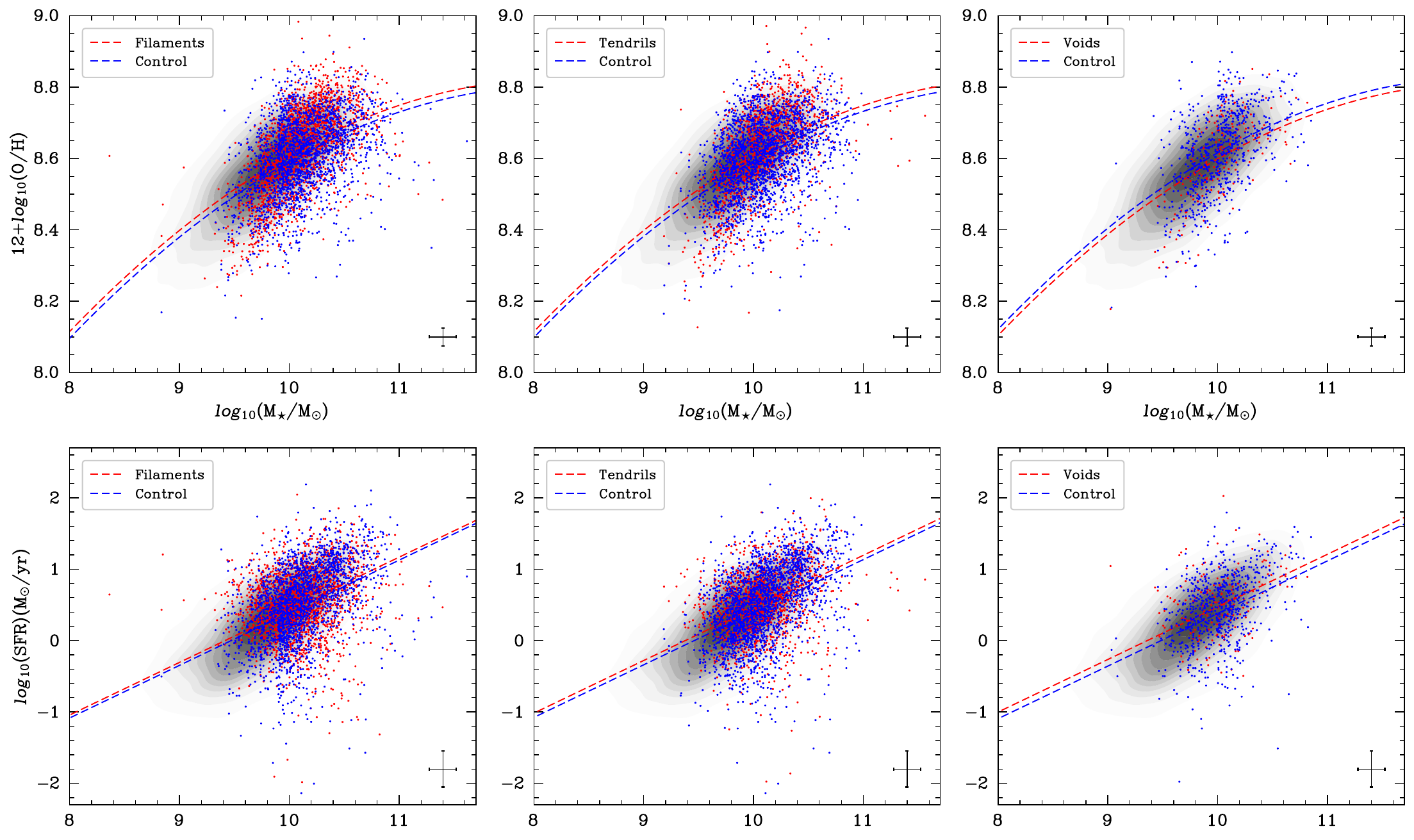}
    \caption{$M_{\star}$--$Z$ (top) and $M_{\star}$--SFR (bottom) relations, using the \citetads{2004MNRAS.348L..59P} and \citetads{2003ApJ...599..971H} calibrations respectively. The galaxies in filaments (left), tendrils galaxies (centre), and voids (right) are marked as red points alongside their corresponding control samples in blue. All SF galaxies are represented in the background as a black density scatter plot. The dashed lines represent the linear regressions calculated by the Bayesian method for each galaxy type (in red) and its corresponding control sample (in blue). Error bars are the median errors of the axes calculated using Monte Carlo propagation.}
    \label{fig: Combined_Plot_B}
\end{figure*}

Figure \ref{fig: Combined_Plot_B} shows the Bayesian fits obtained for the $M_{\star}$--$Z$ and $M_{\star}$--SFR relations. Similarly to the least-squares method,  the $M_{\star}$--$Z$ relation shows small $\Delta Z$ differences in metallicity between the galaxies of each type and their corresponding control samples. In contrast, for the $M_{\star}$--SFR relations the $\Delta$SFR differences  become significant (see Figure \ref{fig: Combined_Plot_B}). Table \ref{tab: Relation_coefB} summarises the differences in metallicity and SFR found, along with their associated errors calculated using the standard deviation of the Markov chains.

\begin{figure}[ht]
    \centering
    \includegraphics[width=\hsize]{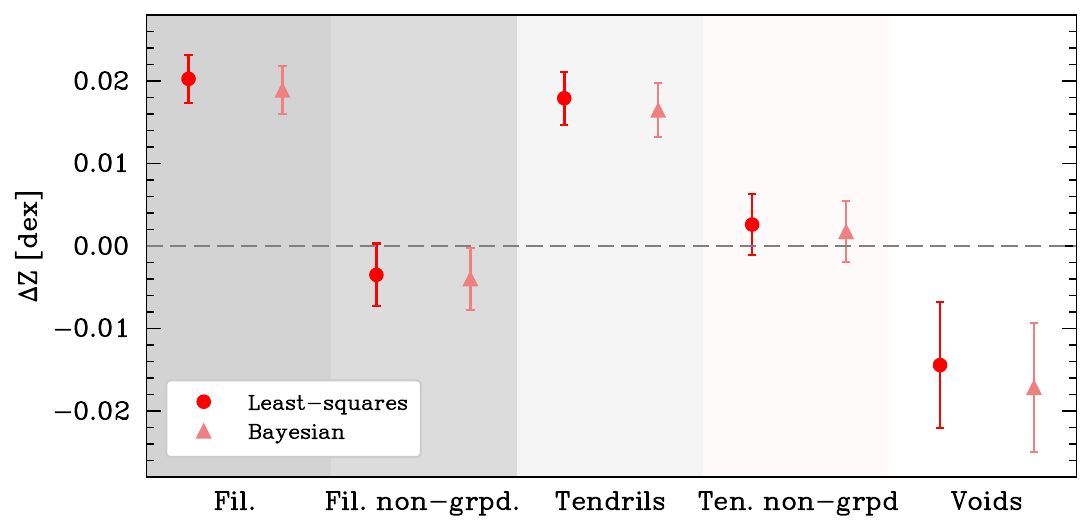}
    \includegraphics[width=\hsize]{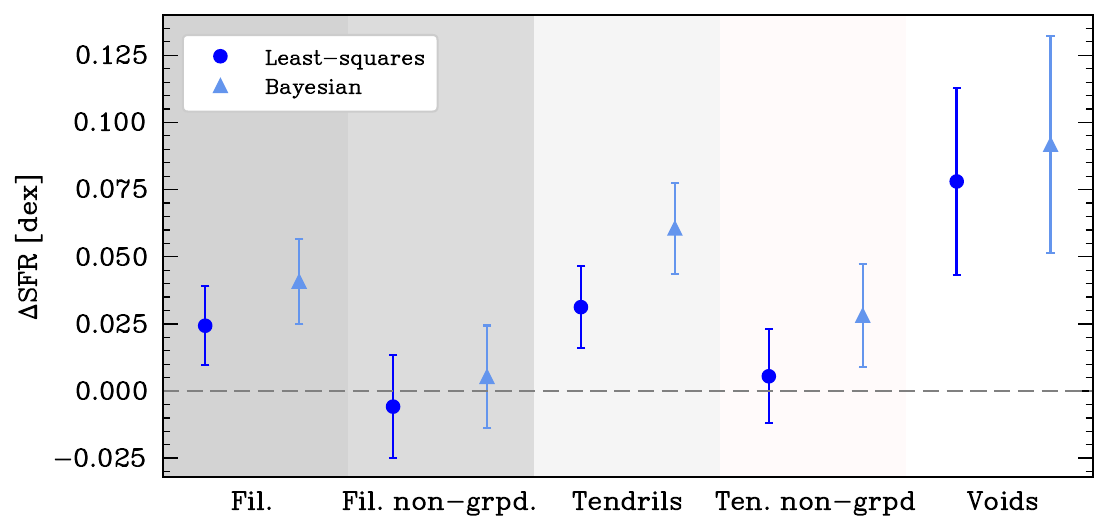}
    \caption{Differences in $Z$ (top) and SFR (bottom) between galaxies in filaments (grouped and non-grouped), tendrils (grouped and non-grouped), and voids and their corresponding control samples, using the least-squares (circle points) and Bayesian (triangle points) methods. The error bars have been calculated using bootstrapping for the least-squares method and 1\,$\sigma$ Markov chains for the Bayesian method.}
    \label{fig: Coefficients_LS_B_Z_SFR}
\end{figure}

Figure \ref{fig: Coefficients_LS_B_Z_SFR} summarises the $\Delta Z$ and $\Delta$SFR results from Tables \ref{tab: Relation_coefLS} and \ref{tab: Relation_coefB} for the least-squares and Bayesian methods.

For the results summarised in Figure \ref{fig: Coefficients_LS_B_Z_SFR}, it is interesting that the Bayesian results are slightly higher compared to the least-squares results for the SFR, and the opposite for Z. These small differences are within the errors, and both are consistent with each other, showing the robustness of our results using either method. Nevertheless, one of the methods could lead to an under- or overestimation. Even so, using the KS and AD tests to compare the $Z$ and SFR distributions of the problem and control samples, we confirm again that the differences found are statistically significant, thus validating the differences found. \\

\begin{figure}[H]
    \centering
    \includegraphics[width=\hsize]{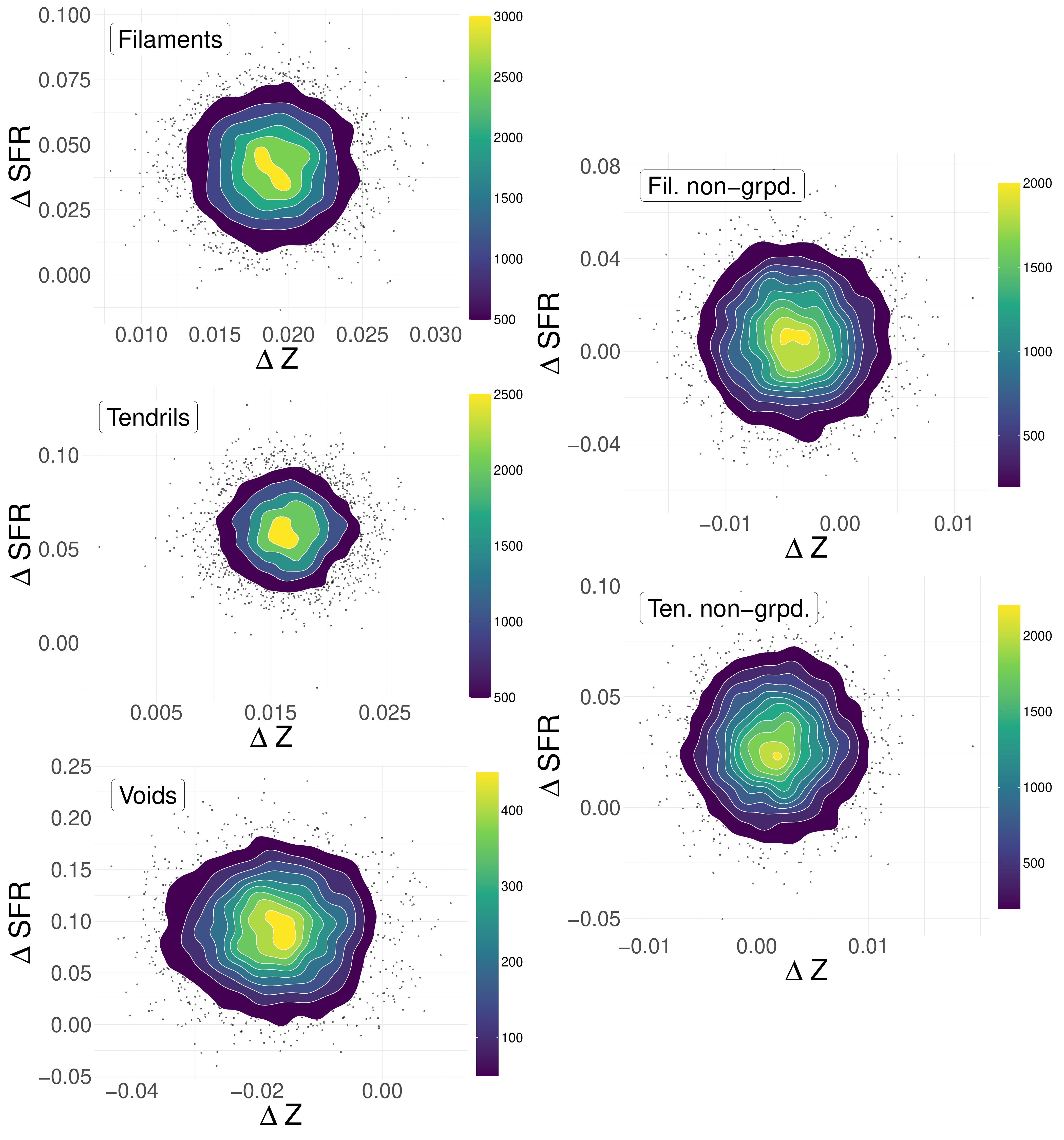}
    \caption{2D Bayesian distributions of the differences in star formation rate ($\Delta$SFR) and metallicity ($\Delta Z$) for galaxies in filaments (grouped: upper left; non-grouped: upper right), tendrils (grouped: middle left; non-grouped: bottom right), and voids (bottom left), compared to their corresponding control samples.}
    \label{fig: DeltaSFR_combined_row}
\end{figure}

Finally, Figure \ref{fig: DeltaSFR_combined_row} shows the Bayesian results for $\Delta Z$ and $\Delta$SFR as full two-dimensional Bayesian distributions. This approach avoids reducing the results to single values with associated errors and provides a two-dimensional probability distribution instead. We note that there is no correlation between the $\Delta Z$ and $\Delta$SFR differences.

\subsection{Non-grouped galaxies} \label{sec: Non-grouped galaxies}

In this section, we analyse galaxies located in filaments and tendrils but not associated with any group, aiming to assess the direct impact of the environment on their properties. To achieve this, we used the GAMA group catalogue (G3C) \citepads{2011MNRAS.416.2640R} to identify filament and tendril galaxies that do not belong to any galaxy group. With this information, we were able to assess whether the evolution of these parameters is more dependent on large-scale structures or local-scale ones.

A control sample, which passed the KS and AD tests (see p-values in Table \ref{tab: p-values-combined}), was created for these non-grouped galaxies in filaments. A total of 1\,435 field galaxies were found compared to 1\,051 non-grouped galaxies in filaments. The same process was applied to non-grouped galaxies in tendrils. In this case, a total of 1\,499 field galaxies were identified, compared to 1\,143 non-grouped galaxies in tendrils.

The values of $\Delta Z$ and $\Delta$SFR obtained for these non-grouped galaxies in filaments and tendrils are also included in Tables \ref{tab: Relation_coefLS} and \ref{tab: Relation_coefB} for the least-squares and Bayesian methods, respectively. In addition, these results are also summarised in Figures \ref{fig: Coefficients_LS_B_Z_SFR} and \ref{fig: DeltaSFR_combined_row}.

\subsection{Morphologies} \label{sec: Morphologies}

Finally, in this section, we analyse the morphology of galaxies in filaments (grouped and non-grouped), tendrils (grouped and non-grouped), and voids with their respective control samples.
Our goal is to see if there are morphological differences that could be attributed to the environment. For this analysis, we used the GAMA Sèrsic catalogue as an indicator of morphology, which provides the Sèrsic index in multiple bands \citepads{2012MNRAS.421.1007K}.

\begin{figure}[ht]
    \centering
    \includegraphics[width=\hsize]{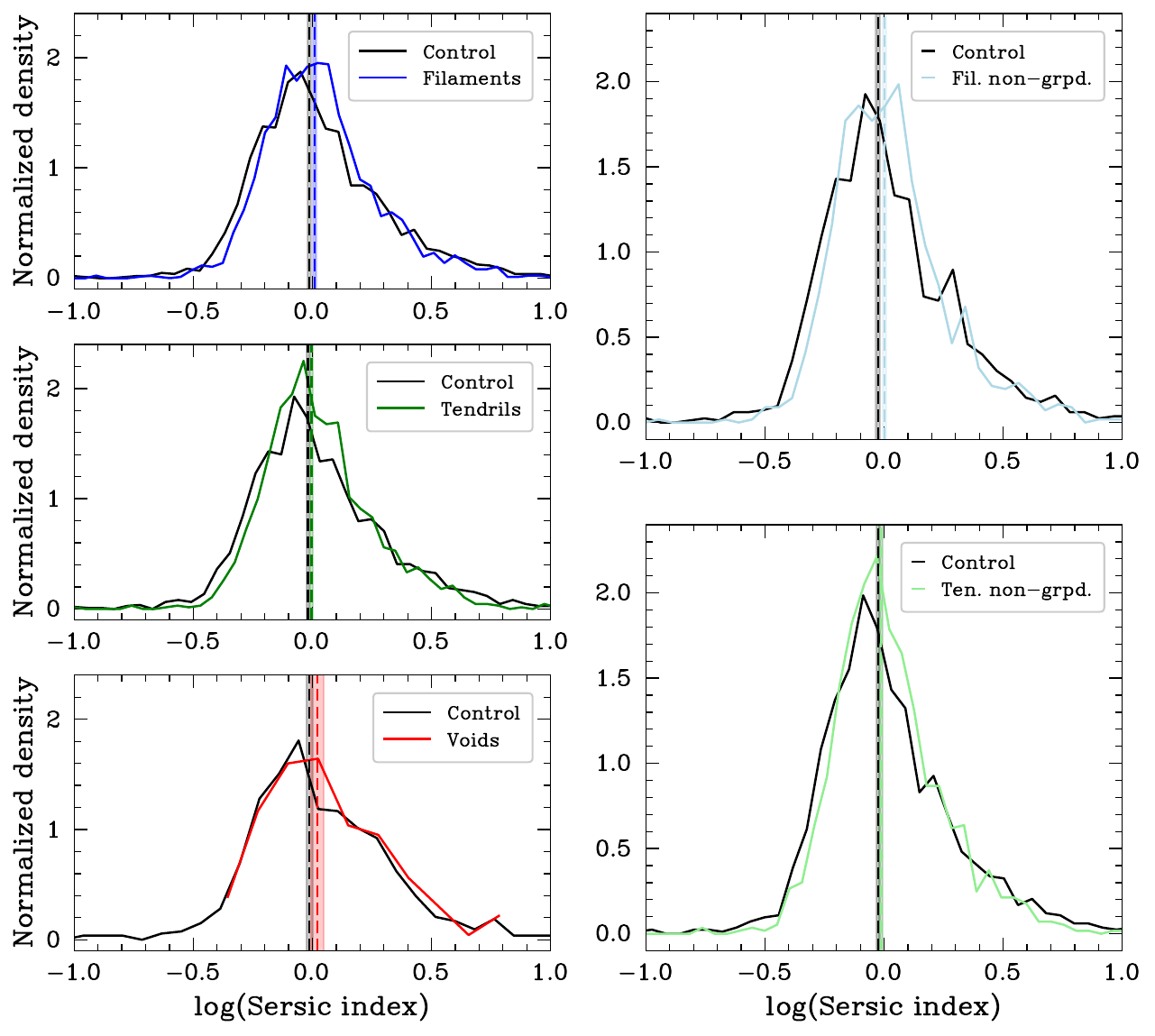}
    \caption{Morphological distribution (logarithm of the Sérsic index in the $r$ band) normalised for the sample of galaxies in filaments (grouped: top left; non-grouped: top right), tendrils (grouped: middle left; non-grouped: bottom right), and voids (bottom left). The horizontal bars indicate the medians, and the shaded areas represent the associated errors.}
    \label{fig: Morphologies}
\end{figure}

Figure \ref{fig: Morphologies} shows the morphology distributions (Sérsic index in the r-band) for each problem sample compared to its control sample.

The errors of the medians (shaded regions in Fig. \ref{fig: Morphologies}) for each distribution overlap with each other. On the other hand, when applying the KS and AD tests to these distributions, they are only satisfied for the void galaxies compared to their control sample. This implies that, although the medians of the distributions are similar, small morphological differences exist. For instance, the shape of the distribution of filaments and tendrils is slightly skewed towards early-type galaxies. 

\section{Discussion} \label{sec: Discussion}

In this work, we describe our analyses of the $M_{\star}$--$Z$ and $M_{\star}$--SFR relations of a selected sample of star-forming galaxies extracted from the GAMA survey (\citeads{2011MNRAS.413..971D}; \citeads{2015MNRAS.452.2087L}). The main objective here was to identify differences in $Z$ and SFR based on the type of galaxy according to the environment (filaments, tendrils, and voids), using generated control samples for each type. To achieve this, Bayesian and classical statistics were used. In this way, we investigated how different extreme environments, from voids to filaments, affect galaxy properties. 

Our analysis revealed variations in $Z$ and SFR across different environments. In particular, galaxies residing in filaments exhibit a higher $\Delta Z$. While their history of undergoing multiple star-forming cycles may contribute to this enhancement (\citeads{2024MNRAS.527.3276I}; \citeads{2024A&A...685A..81G}), it is not clear how the nature of the filamentary environment itself plays a dominant role. It is known that dense environments facilitate the accretion of metal-enriched gas, boosting the metallicity of galaxies in these environments relative to those in less dense regions (\citeads{2006MNRAS.367..139A}; \citeads{2008MNRAS.390..245C}). Furthermore, these phenomena align with the $\Delta$SFR they are associated with, as the feedback environmental processes, combined with the effects of galactic interactions \citepads{2023RAA....23b5016D}, eject the gas that is favourable for incrementing significantly the SFR \citepads{2024ApJ...974..117P}. 
On the other hand, tendril galaxies exhibit a slightly more enhanced $\Delta$SFR and similar $\Delta Z$ in comparison to galaxies in filaments. This suggests that both structures exhibit similar behaviour owing to their small differences.
In contrast, void galaxies exhibit the lowest $\Delta Z$, which has already been observed in other studies, linking this phenomenon to gas accretion by the halo and their SFH \citepads{2023A&A...680A.111D}. However, these galaxies show the highest  $\Delta$SFR from all the studied samples. This can be explained by continuous star formation episodes (\citeads{2016MNRAS.458..394B}; \citeads{2016ApJ...831..118M}; \citeads{2023Natur.619..269D}), i.e., short recurrent star formation bursts arising from cold gas reserves from the halo. Furthermore, the low-density environments in which these objects are immersed prevent other galaxies from stripping their gas, thus allowing for ongoing star formation (\citeads{2022A&A...658A.124D}; \citeads{2024arXiv241102129A}). With all this information, we observe that the $\Delta Z$ and $\Delta$SFR differences obtained (see Figure \ref{fig: Coefficients_LS_B_Z_SFR}) for galaxies in filaments, tendrils and voids are consistent with the results of previous studies (\citeads[e.g.]{2008AJ....135.1877E}; \citeads{2021MNRAS.508.1817S}; \citeads{2024MNRAS.528.4139H}). 

Overall, in filaments and tendrils, interactions between galaxies can trigger star formation. In contrast, in voids, star formation is sustained due to the isolation of galaxies, and the surrounding halo may be accreted, contributing to the enhancement of the SFR. This increase in SFR as one moves towards lower-density environments aligns with the observed decrease in $Z$ in the opposite density direction, which can be attributed to the accretion of enriched gas and interactions for galaxies in filaments and tendrils, and primordial gas accretion from the galaxy's halo for void galaxies. 

On the other hand, galaxies in filaments and tendrils that do not belong to any galaxy group were studied in Section \ref{sec: Non-grouped galaxies}. For these non-grouped galaxies, only minimal differences in $Z$ and SFR were found compared to their control sample of field galaxies (see Figure \ref{fig: Coefficients_LS_B_Z_SFR}). These results suggest that the evolution of ungrouped galaxies, even if they are part of filaments or tendrils, is driven without important external perturbations, leading to a slower and more regulated gas consumption \citepads{2018ApJ...852..142C}. 

The study of these ungrouped galaxies is particularly important in our work for two reasons. First, these galaxies exhibit minimal differences in $Z$ and SFR compared to field galaxies, being largely compatible with them in most cases. Second, these differences are significantly smaller than those observed in the overall population of galaxies in filaments and tendrils, which includes both ungrouped and grouped galaxies. This suggests that the grouped galaxies are responsible for the increase in $\Delta Z$ and $\Delta$SFR. Therefore, the values of $Z$ and SFR are more influenced by the local density  (e.g. group, cluster) in which the galaxy resides, rather than by the large-scale structure (filament, tendril, or void). Consequently, the evolutionary properties of a galaxy are determined primarily by the density of its local environment. This is in agreement with other studies, such as those by \citetads{2004MNRAS.353..713K}, \citetads{2012MNRAS.420.1481V}, \citetads{2013MNRAS.431..167R}, \citetads{2024MNRAS.528.4139H}, and \citetads{2024MNRAS.534.1682O}. Hence, large-scale structure can be understood as a second-order factor that significantly influences the characteristics of galaxies.

In the context of the ‘nature’ versus ‘nurture’ debate, our results suggest that nurture, specifically local density, has a stronger impact on galaxy evolution than mere membership in filaments or tendrils. However, we note that nature, or the intrinsic properties of galaxies, such as internal feedback, gas accretion, or disk instabilities also plays a significant role (\citeads[e.g.]{2004AJ....128.2677T}; \citeads{2010MNRAS.402..282M}; \citeads{2013A&A...550A.115H}; \citeads{2022A&A...661A.105S}), though this aspect is not explored in the present study.
 
Finally, regarding the morphologies studied in Section \ref{sec: Morphologies}, Figure \ref{fig: Morphologies} shows that the errors of the medians for each morphological distribution overlap. When applying the KS and AD tests, they are only satisfied for the void galaxies in comparison to their control sample. Although the medians of the distributions are nearly identical, this suggest that there are small morphological differences in the direction of early types. This has already been analysed in previous studies, such as that by \citetads{2023MNRAS.524.5768P}.

Furthermore, a morphologically driven quenching can be ruled out for SF galaxies, as the Sérsic index distribution of galaxies in filaments, tendrils, and voids is similar to that of their corresponding control samples (\citeads[e.g.]{2017ApJ...846L...4R}; \citeads{2017MNRAS.464..666B}; \citeads{2022MNRAS.517.4575S}). Lastly, it is worth noting that there is no correlation between the $\Delta Z$ and $\Delta$SFR differences (see Figure \ref{fig: DeltaSFR_combined_row}). 

\section{Conclusions} \label{sec: Conclusions}

This study shows that the environment affects galaxy metallicity and SFR. The main conclusions are as follows:

\begin{itemize}

    \item Galaxies in filaments exhibit higher metallicity and lower star formation rates, due to their enriched environments and ongoing interactions.

    \vspace{0.10cm}
    
    \item Void galaxies have the lowest metallicity but show high star formation rates driven by recurrent bursts fuelled by halo cold gas reserves and a lack of gas stripping.

    \vspace{0.10cm}

    \item The $\Delta Z$ and $\Delta$SFR differences found suggest that the local environment, rather than large-scale membership, has a greater influence on galaxies’ evolutionary properties.

    \vspace{0.10cm}

    \item A morphologically driven quenching can be ruled out based on the similarity of Sérsic index distributions. 
  
\end{itemize}

Metallicity and SFR are strongly linked to environment, underlining the need for new surveys to better assess how structural context influences galaxy evolution.

\begin{acknowledgements} \label{sec: Acknowledgments}

    We thank the referee for their thoughtful comments that improved the manuscript.
    
    This work has been supported by grant PID2021-123417OB-I00 and PCI2022-135023-2 funded by the Spanish Ministry of Science and Innovation. M.A.L.L. acknowledge support from the Ramón y Cajal programme funded by the Spanish Government  RYC2020-029354-I.
    
    In addition to other packages mentioned, we made use of: \texttt{astropy} \citepads{2013A&A...558A..33A}, \texttt{matplotlib} \citepads{2007CSE.....9...90H}, \texttt{numpy} \citepads{2020Natur.585..357H}, and \texttt{smplotlib} \citepads{2023zndo...7966831L}; as well as the following R packages: \texttt{ggplot2} \citep{Wickham2016-ic}, \texttt{patchwork} \citep{patchwork}, and \texttt{propagate} \citep{propagate}. 
    
    GAMA is a joint European-Australasian project based around a spectroscopic campaign using the Anglo-Australian Telescope. The GAMA input catalogue is based on data from the Sloan Digital Sky Survey and the UKIRT Infrared Deep Sky Survey. Complementary imaging of the GAMA regions is being obtained by a number of independent survey programmes, including GALEX MIS, VST KIDS, VISTA VIKING, WISE, Herschel-ATLAS, GMRT and ASKAP, providing UV to radio coverage. GAMA is funded by the STFC (UK), the ARC (Australia), the AAO and the Participating Institutions. The GAMA website is \url{http://www.gama-survey.org/}.

\end{acknowledgements}

\bibliography{bibliography} 

\end{document}